\documentclass[preprint,12pt]{iopart}
\usepackage{amssymb}
\expandafter\let\csname equation*\endcsname\relax
\expandafter\let\csname endequation*\endcsname\relax
\usepackage{amsmath}
\usepackage{booktabs}
\usepackage{graphicx}
\usepackage{floatrow}
\usepackage[caption=false]{subfig}
\usepackage{ulem,xcolor}
\graphicspath{{figures/}}

\newcommand{\tc}[1]{#1}

\begin{document}

\title[A treecode to simulate dust-plasma interactions]{A treecode to simulate dust-plasma interactions$^\dagger$}
\footnotetext{Substantial portions of this paper \tc{are adapted} from chapters 4 and 5 of the first author's recent PhD dissertation.}
\author{D M Thomas$^1$ and J T Holgate$^1$\footnote{Corresponding author.}}
\address{$^1$ Blackett Laboratory, Imperial College London, London SW7 2BW, United Kingdom}
\eads{\mailto{dmt107@imperial.ac.uk}, \mailto{j.holgate14@imperial.ac.uk}}

\begin{abstract}
The interaction of a small object with surrounding plasma is an area of plasma-physics research with a multitude of applications.
%Hitherto, macroscopic particle-in-cell (PIC) codes have dominated the computational study of these interactions.
This paper introduces the plasma octree code \texttt{pot}, a microscopic simulator of a spheroidal dust grain in a plasma.
\texttt{pot} uses the Barnes-Hut treecode algorithm to perform $N$-body simulations of electrons and ions in the vicinity of a chargeable spheroid, employing also the Boris particle-motion integrator and Hutchinson's reinjection algorithm from SCEPTIC; a description of the implementation of all three algorithms is provided.
We present results from \texttt{pot} simulations of the charging of spheres in magnetized plasmas, and of spheroids in unmagnetized plasmas.
The results call into question the validity of using the Boltzmann relation in hybrid PIC codes.

\end{abstract}


\pacs{  % https://publishing.aip.org/publishing/pacs/pacs-reg50
55.65.Cc  % "Particle orbit and trajectory"
55.65.Yy  % "Molecular dynamics methods"
55.27.Lw  % "Dusty or complex plasmas; plasma crystals"
}

\submitto{\PPCF}

%% main text
\section{Introduction}
\label{sec:Intro}
The study of dusty plasmas is concerned with objects, usually on the micro- or nano-scale, immersed in a hot ionized gas known as a plasma. These objects, referred to as dust grains, may be either solid or liquid and are ubiquitous in plasmas. As such, the instances and applications of dusty plasmas are too numerous to elaborate on fully here; they include interstellar dust, planetary rings, noctilucent clouds, plasma spraying, contamination in semiconductor processing plasmas and impurities in magnetic confinement fusion devices, to list but a few \cite{Shukla2002}.

The collective behaviour of a pure plasma is highly complex, and depends on the interactions between vast numbers of individual ions and electrons. This complexity is increased further by the inclusion of dust grains; not only do they represent an additional charged species, but they act as sources and sinks for electrons and ions. Therefore the charge on the dust may fluctuate \cite{Thomas2013}, but this charge depends additionally on non-plasma processes such as thermionic, field-induced and photonic emission of electrons \cite{Sodha2014}. The shape and size of dust is also variable, as they can grow through aggregation \cite{Mohideen1998} or shrink through evaporation and violent processes such as electrostatic breakup \cite{Coppins2010}. The co-dependence of these processes, and many others, in a dusty plasma has led to their alternative name of ``complex plasmas'', and is manifest in surprising phenomena such as the self-organization of dust grains into crystal-like structures \tc{\cite{Miloshevsky2012a}}. Although some approximate analytic theories exist to describe fundamental processes in a dusty plasma, the inherent complexity of these systems necessitates computer simulations to resolve their full detail.

As an illustration of the difficulty in modelling dust-plasma interactions, consider the most fundamental process in a dusty plasma: the charging of grains by ion and electron currents drawn from the plasma. The most widely used theory to describe these currents is the orbit-motion-limited (OML) theory \cite{Allen1992}, which gives simple algebraic expressions for the currents but assumes a small, spherical grain, no potential barrier to ions reaching the grain, a stationary and Maxwellian plasma far from the grain, no applied electric or magnetic fields, no trapped ions, no ionization, and no collisions between particles. The OML assumptions have provoked some criticism \cite{Allen2000}, but for small grains the theory works remarkably, even surprisingly \cite{Chen2009}, well. OML theory has been extended to more realistic cases, such as the shifted orbit-motion-limited (SOML) theory, which applies to drifting Maxwellian plasmas \cite{Willis2012}. The inclusion of ion collisions with neutrals, which trap ions in orbits around the dust and increase the ion current to the grain, have also been studied \cite{Lampe2001}. A more controversial extension to OML theory has been the addition of magnetic fields by assuming that only the electrons become magnetized \cite{Tsytovich2003}; that is to say that the electrons follow helical trajectories due to the magnetic Lorentz force, while the heavier ions are unaffected on the scale of the dust.

While these extended theories improve on the OML model, they still omit several important features of real plasmas. Similar complexity is faced in all other aspects of dust-plasma interactions; for example in calculating the drag force of the plasma on the dust, the plasma response to the dust, and wave propagation in dusty plasmas. Only computer simulations can provide complete models of the dust-plasma interactions, and to this end several particle-in cell (PIC) codes have been developed \tc{\cite{Hutchinson2002a, Matyash2006, Miloch2007, Miloshevsky2012b, Anuar2013, Delzanno2013, Lange2016}}. The most widely used of these, SCEPTIC, shows excellent agreement with the OML and SOML models in the appropriate limits \cite{Hutchinson2005}. However, PIC codes incorporate only some of the microscopic detail of the plasma, because the fields are interpolated from grid points and the inter-particle forces are underestimated within cells. Collisions between particles must therefore be artificially imposed on the simulation to be included at all, despite the fact that they can be crucial to many aspects of the dust-plasma interaction \cite{Lampe2001}. Furthermore, hybrid codes such as SCEPTIC employ the Boltzmann relation for electrons, which may be invalid when a magnetic field is present \cite{Allen2008}, and which dispenses completely with microscopic information about the electrons. A sceptic might therefore suppose that the analytic theories and PIC codes agree only because they share systematic biases arising from the details they both omitted.

To preempt this criticism one could calculate the motion of every single particle in the plasma in order to maximize the faithfulness of a simulation. Insofar as such a simulation successfully approximated the motion of every ion and electron, it would necessarily produce results like those of a real plasma. However, given a plasma of $N$ particles, computing the field felt by one particle requires iteration over the remaining $N-1$ particles, and repeating this for all of the particles results in a computational time proportional to $N(N-1)$. The runtime of an exact simulation is therefore $\mathcal{O}(N^2)$, which becomes prohibitively large as $N$ is increased to that required for a realistic simulation.

The treecode algorithm developed by Barnes and Hut for galactic simulations allows this formidable runtime cost to be avoided by calculating approximate, rather than exact, values for the field in $\mathcal{O}(N\log N)$ time \cite{Barnes1986}. \tc{This algorithm has already demonstrated its utility in simulations of laser-plasma interactions \cite{Gibbons2010} and of some 1D and 2D low-temperature plasma applications \cite{Christleib2006}}. This paper describes the development of the fully microscopic plasma octree code \texttt{pot}, which implements the Barnes-Hut algorithm for a plasma in the vicinity of a dust grain. \tc{The term ``octree'' here refers to the algorithm's eightfold splits of the 3D space within the simulation.}

\texttt{pot} has several novel features to commend it to the computational physicist: it is the first \tc{3D} implementation of the Barnes-Hut algorithm in a low-temperature \tc{or dusty} plasma environment, it provides a rare example of the Boris particle-motion integration scheme \cite{Boris1970} outside particle-in-cell (PIC) codes, and it contains the first successful re-implementation of Hutchinson's particle-injection algorithm beyond SCEPTIC \cite{Hutchinson2003}.

An overview of the implementation and scope of \texttt{pot} is given in section \ref{sec:Overview}. It would be tedious to describe the source code of \texttt{pot} in its entirety but there are three algorithms which, being vital to \texttt{pot}'s successful implementation, deserve elucidation. These are the Boris particle-motion integrator \cite{Boris1970}, the Barnes-Hut treecode \cite{Barnes1986}, and Hutchinson's reinjection algorithm \cite{Hutchinson2003}, and section \ref{sec:Algorithms} provides their specifications. However it is not enough to have just a computer program which simulates the dust-plasma interaction; one has to have some grounds for trusting its output, and section \ref{sec:Results} gives details of some tests of \texttt{pot} to check that it gives physically realistic results. In particular the charging behaviour of \texttt{pot} is compared against the predictions of the OML, SOML, and magnetized-electron theories, and of SCEPTIC. The first new results from \texttt{pot}, for the charging of spherical grains in a strong magnetic field and the charging of non-spherical grains in unmagnetized plasmas, are also presented. Section \ref{sec:Summary} provides a brief summary of this work, with a look towards the future applications of \texttt{pot}.

\section{The plasma octree code \texttt{pot}}
\label{sec:Overview}
The plasma octree code \texttt{pot}, in its present form, simulates a lone collecting spheroid (the dust grain) in a spherical region of wholly ionized plasma, with the user able to choose the size of both the collecting spheroid and the simulation region. The plasma consists solely of $\lceil N/2 \rceil$ electrons and $\lfloor N/2 \rfloor$ ions of one pre-defined species, where $N$ is \tc{selected by the user subject to runtime and memory constraints.} The program simulates the plasma by approximately solving the trajectories of every particle using the Boris integrator \cite{Boris1970}, the precise specification of which is given in section \ref{sec:Boris}. The electrons and ions are modelled as classical, non-relativistic point charges of constant mass \tc{while the dust grain is taken to be a perfectly conducting spheroid. These particles} interact with each other through their electrostatic fields. Coulomb collisions are therefore inherent and need not be artificially imposed on the simulation.

The user determines the values of the plasma parameters: \texttt{pot} accepts the electron and ion temperatures as command-line arguments while the electron and ion masses and ion charge may be adjusted by changing compile-time constants in \texttt{pot}'s source code. The plasma density cannot be set directly, but is implied by the user's choice of $N$ and the simulation domain's size.

To save processing time, \texttt{pot} assumes that the time-varying magnetic interactions between particles due to their motion are negligible compared to their electrostatic interactions; this is consistent with the fact that \texttt{pot} simulates non-relativistic plasmas. However, the user is able to impose an arbitrary space- and time-independent magnetic field on the plasma. Each simulated particle experiences the usual Lorentz force, with the time-varying electric field computed from the system's charge distribution \textit{via} the treecode algorithm as specified in section \ref{sec:Barnes-Hut}.

Each simulation begins with the electrons and ions distributed randomly with their positions sampled from a uniform distribution and their velocities sampled from a drifting Maxwell-Boltzmann distribution with a user-specified flow speed. The dust begins with no charge but rapidly acquires it through the collection of electrons and ions, the trajectories of which are interpolated between timesteps to ensure accurate collection. Ions or electrons which leave the simulation, either by colliding with the dust grain or by breaching the simulation region's boundary, are reinjected at a random point on the simulation region's boundary \tc{(\textit{q}.\textit{v}.\ section \ref{sec:Hutchinson})}. One might expect that the particles can simply be reinjected according to the Maxwell-Boltzmann distribution; however, this fails to account for the geometry of the simulation domain. Any smooth, contiguous region of the simulation's boundary faces in a particular direction, and this anisotropy causes the velocity distribution of particles entering the domain to differ from a Maxwell-Boltzmann distribution. The reinjection algorithm developed by Hutchinson for SCEPTIC takes this effect into account \cite{Hutchinson2003}, but it lacks a comprehensive written exposition. A detailed review of this method, and the differences in its implementation between \texttt{pot} and SCEPTIC, is provided in section \ref{sec:Hutchinson}.

\begin{figure}
%\floatbox[{\capbeside\thisfloatsetup{capbesideposition={right, center},capbesidewidth=0.4\textwidth}}]{figure}[\FBwidth]
{\caption{An example of \texttt{pot}'s graphical user interface (GUI) for a simulation containing 2500 (blue) electrons, 2500 (red) ions and a single (green) dust grain with an applied magnetic field of $1$ T\tc{; full-scale simulations have been performed with 150000 particles in total}. The particles' helical trajectories and the dust grain are shown to scale while the ions and electrons, being microscopic and invisible on this scale, are represented by spheres much larger than their physical size. The GUI can be initialized by compiling \texttt{pot} using a flag definable in \texttt{pot}'s source code. It has been particularly useful in testing for sensible particle collection and reinjection and for ensuring steady particle gyro-orbits.}\label{fig:GUI}}
{\includegraphics[width=0.54\textwidth]{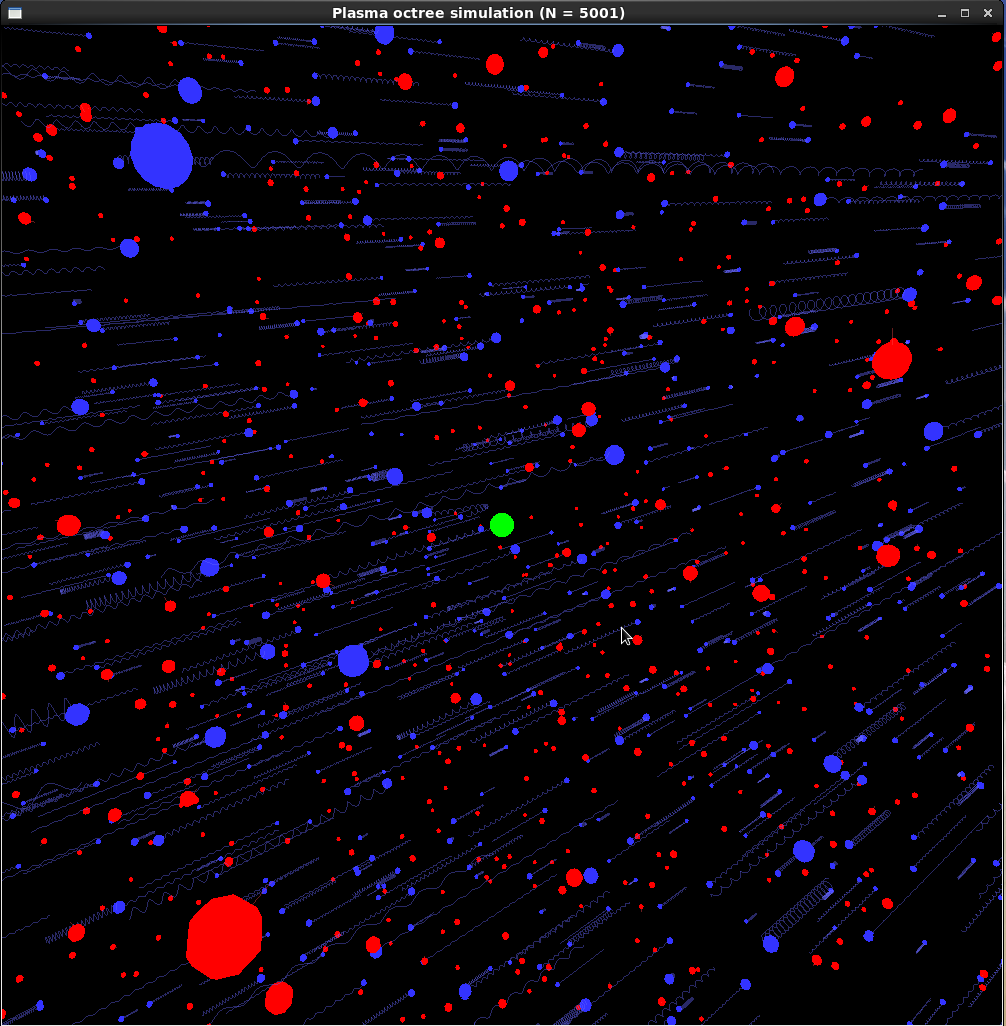}}
\end{figure}

\texttt{pot} is a parallel program, written in C, which uses the Message Passing Interface (MPI) to divide tasks across multiple processes.
It can be compiled to display the simulated particles' motion and trajectories (figure \ref{fig:GUI}), live, using the OpenGL graphical library, which has proved a valuable visualization and debugging tool.
The program is available online at \texttt{https://github.com/drewthomas/pot}.

\section{Core algorithms}
\label{sec:Algorithms}
The successful operation of \texttt{pot} relies on several interlocking algorithms. The Boris particle-motion integrator \cite{Boris1970}, the Barnes-Hut treecode \cite{Barnes1986} and Hutchinson's particle-reinjection algorithm \cite{Hutchinson2003} are particularly vital to \texttt{pot}. Because it can be inconvenient to locate lucid, precise specifications of these algorithms, and because their implementation in \texttt{pot} may differ from elsewhere, the following subsections describe their implementation in \texttt{pot}.

\subsection{Boris particle-motion integrator}
\label{sec:Boris}
The equation of motion for a non-relativistic plasma particle is simply Newton's second law with the Lorentz force substituted into it,
\begin{equation}
\left.
\begin{aligned}
 \frac{\textrm{d}\boldsymbol{r}(t)}{\textrm{d}t} &= \boldsymbol{v}(t) \\
 \frac{\textrm{d}\boldsymbol{v}(t)}{\textrm{d}t} &= \frac{q}{m}[\boldsymbol{E}(\boldsymbol{r}(t),t)+\boldsymbol{v}(t) \times \boldsymbol{B}], \quad
\end{aligned}
\right\}
\end{equation}
where $\boldsymbol{r}(t)$ and $\boldsymbol{v}(t)$ are the time-varying position and velocity of the particle, which has mass $m$ and charge $q$ and is subjected to the electric and magnetic fields $\boldsymbol{E}(\boldsymbol{r}(t),t)$ and $\boldsymbol{B}$. It would be impossible to solve these equations analytically for every particle, so a range of integrators have been devised which, given the values $\boldsymbol{r}(t_0)$ and $\boldsymbol{v}(t_0)$, progress the simulation through a timestep of length $\delta t$ to give the updated values $\boldsymbol{r}(t_0 + \delta t)$ and $\boldsymbol{v}(t_0 + \delta t)$. The trajectories of all the particles can be evaluated over time by executing the integrator iteratively. The time step must be small enough to resolve the motion of the particles, so is constrained by the following conditions.
\begin{enumerate}
\item Particles with temperature $T$, and hence average thermal velocity $v_{th}=(k_BT/m)^{1/2}$, must not be able to traverse the grain, of radius $a$, in one time step, which imposes the requirement $\delta t \ll a / v_{th}$. \tc{For a non-spherical grain the appropriate value of $a$ is the smallest radius of curvature at any point on the surface, for example at the needle-like tips of an elongated prolate spheroid.}
\item As the simulation must resolve collective motion of the particles, the timestep must be shorter than the period of an electron plasma oscillation. Therefore $\delta t \ll \omega_{pe}^{-1}$, where $\omega_{pe}=(n_e e^2/m_e\epsilon_0)^{1/2}$ for a plasma with electron density $n_e$.
\item The integrator must resolve the gyromotion of a particle around the magnetic field lines, which occurs with frequency $\omega_g=|q|B/m$ and hence $\delta t \ll \omega_g^{-1}$. 
\end{enumerate}
$\delta t$ is left to the \texttt{pot} user's discretion, who must keep these constraints in mind. \tc{Criterion (i) is generally the most stringent. For example, taking the default parameters listed in table \ref{tab:defaults} gives $\delta t \ll 43.3$ ps and $\delta t \ll 6.66$ ns for the first two conditions respectively. The default timestep of $1$ ps is therefore appropriate and this additionally satisfies condition (iii) provided $B < 5.69$ T or, as defined in equation \ref{eq:beta-i-defn}, $\beta_i<0.81$.} The ions, being much heavier and slower than the electrons, may have a \tc{longer} integration timestep than the electrons, so the positions and velocities of the ions are updated every $(T_em_i/T_im_e)^{1/2}$ steps (rounded down to the nearest integer) while the electrons are updated at every step.

\texttt{pot} implements the Boris integrator according to the concise specification of Patacchini and Hutchinson \cite{Patacchini2009}, rather than the specification in Boris's original paper \cite{Boris1970}, as
\begin{equation}
\left.
\begin{aligned}
 \boldsymbol{r}(t_0+\delta t/2) = & \ \boldsymbol{r}(t_0) + \frac{\delta t}{2} \boldsymbol{v}(t_0) \\
 \boldsymbol{v}(t_0+\delta t) = & \ \mathbf{R}_{\boldsymbol{\delta \varphi}} \left[ \boldsymbol{v}(t_0) + \frac{\delta t}{2} \frac{q\boldsymbol{E}(\boldsymbol{r}(t_0+\delta t/2),t_0+\delta t/2)}{2m} \right] \\
& \ + \frac{\delta t}{2} \frac{q\boldsymbol{E}(\boldsymbol{r}(t_0+\delta t/2),t_0+\delta t/2)}{2m} \\
 \boldsymbol{r}(t_0+\delta t) = & \ \boldsymbol{r}(t_0 + \delta t /2) + \frac{\delta t}{2} \boldsymbol{v}(t_0 + \delta t)
\end{aligned}
\right\}
\end{equation}
where $\mathbf{R}_{\boldsymbol{\delta \varphi}}$ is an operator representing a rotation with magnitude and axis defined by the characteristic vector
\begin{equation}
\boldsymbol{\delta \varphi}
\equiv \frac{2 \boldsymbol{B}}{|\boldsymbol{B}|} \tan^{-1} \left( \frac{\delta t}{2}\omega_g \right).
\end{equation}
This scheme physically represents a drift followed by a kick, followed by another drift. The kick step itself comprises three parts: a half-step of acceleration due to $\boldsymbol{E}$, a full step of gyrorotation around $\boldsymbol{B}$, and another half-step of acceleration by $\boldsymbol{E}$. A shrewd feature of the algorithm is that the electric field must be evaluated only once, in the middle of a time step, which reduces runtime and gives the method second-order accuracy. The time symmetry of the drift-kick-drift and $\boldsymbol{E}$-$\boldsymbol{B}$-$\boldsymbol{E}$ sequences also gives the Boris algorithm time reversibility, so that the error in the total energy of the simulation remains bounded indefinitely \cite{Qin2013}.

\subsection{Barnes-Hut treecode}
\label{sec:Barnes-Hut}
Although the Boris integrator provides a method for updating particle positions, it does not specify how to evaluate the electric field required for the kick step. Evaluating the electric field by applying Coulomb's law to every particle in turn leads to $\mathcal{O}(N^2)$ runtime; the Barnes-Hut treecode algorithm cuts this to $\mathcal{O}(N \log N)$. The algorithm achieves this by replacing distant clusters of particles with a single charge, and computing the electric field due to this effective charge rather than each individual particle. This reduces the number of interactions and accelerates the simulation. However the interactions with nearby particles must still be calculated with high precision, so these must still be treated individually; only long-range interactions, being weak and tending to cancel out, can be clustered. The treecode provides a method of formally defining particle clusters, but avoids completely recomputing them at every position where the electric field is being evaluated.

The algorithm begins by dividing the simulation domain into $2^D$ cells by splitting it in half along each of its $D$ dimensions. If a cell contains more than one particle then it is split again into $2^D$ smaller cells, and this process is repeated until each of the smallest cells contains at most one particle. The resulting hierarchy of cells has a natural representation as a tree, whence the treecode method's name. Specifically, \texttt{pot}, being a 3D simulator, splits each cell into eight cubic cells which motivates the term ``octree''. The simulation visits every cell of every size, recording each cell's total charge and its centre of charge
\begin{equation}
 \boldsymbol{r}_{c} = \frac{\sum_j |Q_j| \boldsymbol{r}_j}{\sum_j |Q_j|}
\end{equation}
where the sums are over all of the particles in the cell, the denominator is nonzero, and $Q_j$ and $\boldsymbol{r}_j$ are the charge and position of particle $j$. The simulation then refers to to the tree to rapidly define clusters for calculating the electric field felt by each particle.

It does this by applying an opening angle criterion to each cell, which decides whether the cell is far enough away that the precise charge distribution of its contents can be replaced by its total charge located at its centre of charge, as already calculated by the treecode. If a particle is at distance $d$ from a cell's centre of charge, and that cell has sides of length $l$, then this criterion is simply whether $l/d < \theta$, where $\theta$ is a fixed opening angle parameter. The default value of $\theta$ is $1$ in \texttt{pot}, but this can be changed by the user to modify the severity of the clustering approximation; for large values of $\theta$ the treecode will group charges into a small number of large clusters. The simulator steps down each branch of the tree hierarchy until the opening angle criterion is satisfied, at which point no smaller sub-cells need to be considered. The number of cells visited in order to estimate the field at a particle is of order $\log N$, so the time to estimate the field at all $N$ particles is $\mathcal{O}(N\log N)$. Building the tree also requires $\mathcal{O}(N\log N)$ time, but it is only built once for all particles so this does not affect the asymptotic $N$ dependence of the algorithm.

The treecode may be modified to improve the accuracy of the approximation; one such modification is to include the dipole moment of each cell in the calculation of the field felt by a particle, as suggested in Barnes and Hut's original paper.
\tc{(Indeed, another treecode-like computational method, the fast multipole method, can perform $N$-body simulations in $\mathcal{O}(N)$ time by including such higher-order multipole expansions, but the method calls for implementing a more involved algorithm \cite{Greengard1994}.)}
This does not impose particularly onerous additional runtime costs as the number of interactions remains the same. \texttt{pot}'s implementation of the treecode algorithm offers the compile-time option of including cells' electric dipole moments to compute particle-cluster interactions.

\subsection{Hutchinson's particle-reinjection algorithm}
\label{sec:Hutchinson}
When an electron or ion is collected on the dust or leaves the simulation domain it must be reinjected into the simulation in order to conserve the particle number density. However, as previously mentioned, simply reinjecting the particle at a random point on the simulation domain boundary with a velocity sampled from a Maxwell-Boltzmann distribution is insufficient to maintain the desired particle velocity distribution of the entire plasma; \tc{tests of this naive reinjection method during \texttt{pot}'s early development} showed that it \tc{made} the \tc{plasma's} equilibrium \tc{velocity} distribution leptokurtic\tc{, with a} temperature roughly a third \tc{less} than the injection distribution's temperature. \tc{This arose from the combination of the simulation domain losing fast-moving particles faster than slow-moving particles, and the geometric fact that a small surface element of the simulation boundary presents a smaller cross-section to particles approaching it nearly tangentially than to particles approaching it nearly perpendicularly (thus particles passing through a boundary surface element are relatively likely to have their velocity aligned towards the surface element's normal).}

The reinjection algorithm must therefore account for the shape of the simulation domain boundary and the hypothetical motion of particles outside it, so that the reinjected particles have velocities as if sampled from \tc{an undistorted} Maxwell-Boltzmann distribution at infinity.
Hutchinson, when designing SCEPTIC, solved this problem for a spherical domain and \texttt{pot} implement\tc{s} the published description of his reinjection algorithm \cite{Hutchinson2003}. Hutchinson's exposition is not comprehensive so, as well as paraphrasing it, the description given here aims to fill some gaps for the convenience of anyone wishing to understand the method's implementation in SCEPTIC and \texttt{pot}, or anyone wishing to develop another simulation using this method. The notation in this subsection follows Hutchinson by writing a particle's velocity at infinity as $\boldsymbol{u}$, the plasma's flow velocity at infinity as $\boldsymbol{U}$, their speeds as $u$ and $U$ respectively, and the cosine of the angle between $\boldsymbol{u}$ and $\boldsymbol{U}$ as $c$.

Making the assumptions that the potential $\phi(\boldsymbol{r})$ is \tc{spherically} symmetric and contains no barriers outside the simulation domain, Hutchinson writes a formula for the flux into the spherical simulation domain ``in the velocity element $\textrm{d} u$ from a distant solid angle element" \cite[p. 1482]{Hutchinson2003}. The only anisotropy in the distant velocity distribution is due to the plasma flow, so this flux may be written in terms of $u$, $U$ and $c$.

Hutchinson, using his expression for the differential flux, then deduces cumulative distribution functions (CDFs) of the probability distributions of $c$ and $u$ for a particle entering the simulation domain \cite[p. 1483]{Hutchinson2003}. $c$'s CDF depends only on $c$, $u$ and $U$, and once $u$ has been sampled it is trivial to generate $c$ values by inverse transform sampling.
%Hutchinson's formula for $c$'s CDF fails when $U=0$ but it can be deduced, from the spherical symmetry that exists without flow, that $c$ has a uniform distribution over $[-1,1]$ in this case.
The CDF for $u$ is more complicated, depending on the normalized electric potential $\chi_b=q\phi(r_b)/(k_BT)$ at the simulation boundary $r=r_b$. Nonetheless, this CDF may also be inverted numerically, and $u$ sampled, by interpolation (as SCEPTIC does) or Newton-Raphson iteration (as \texttt{pot} does).

Once $u$ and $c$ have been sampled, the next stage is to sample a value for the particle's distant impact parameter $b$. A particle entering the simulation domain with given values of $u$ and $c$ must have had a $b$ value between $0$ and $b_{\textrm{max}}=r_b(1-\chi_b/u^2)^{1/2}$, so $b^2$ is sampled uniformly from the range $[0,b_{\textrm{max}}^2]$.

Having sampled $u$ and $b$, the algorithm must determine where the particle enters the simulation. This is achieved by first calculating ``the angle $\alpha$ in the plane of impact between the position of impact [where the particle reaches the simulation's boundary] and the direction of the [initial particle] position at infinity" by evaluating the orbit integral
\begin{equation}
 \alpha \equiv \int_0^1 \left[ \frac{r_b^2}{b^2} \left( 1 - \frac{\chi(r_b/r)}{u^2} \right) - \left( \frac{r_b}{r} \right)^2 \right]^{-\frac{1}{2}} \textrm{d} \left( \frac{r_b}{r} \right)
\end{equation}
where $\chi(r_b/r)$ denotes the dimensionless potential $q\phi(r)/(k_BT)$ \cite[p. 1484]{Hutchinson2003}. The solution of this integral requires knowledge of $\chi(r_b/r)$ outside the simulation domain, in the range $0 \leq r_b/r \leq 1$. \texttt{pot} assumes $\chi(r_b/r)$ to be the electron-only form of the Debye-H\"{u}ckel potential \cite{Hutchinson2002b} and solves for $\alpha$ with an adaptive Simpson's rule, while SCEPTIC uses a more elaborate version of the Debye-H\"{u}ckel profile, which incorporates ion depletion due to absorption on the dust, and evaluates the integral by the trapezium rule \cite[p. 1481 \& 1484]{Hutchinson2003}.

These calculations do not fully determine the injected particle's initial position and velocity; it remains to ``[c]hoose the ignorable angles of the position and impact parameter from $0$ to $2\pi$" \cite[p. 1484]{Hutchinson2003}, although Hutchinson does not provide a concrete procedure to accomplish this. \texttt{pot}'s procedure is instead described here.

The vector $\boldsymbol{u}$ is determined first. In \texttt{pot} the plasma flow $\boldsymbol{U}$ is always in the $\boldsymbol{\hat{x}}$-direction, so $u_x=\boldsymbol{u} \cdot \boldsymbol{\hat{x}} = uc$. The magnitude of $\boldsymbol{u}$ perpendicular to $\boldsymbol{\hat{x}}$ is therefore $u(1-c^2)^{1/2}$, which is oriented in the $\boldsymbol{\hat{y}}$-$\boldsymbol{\hat{z}}$ plane with a polar angle chosen randomly from a uniform distribution over the range $0$ to $2\pi$. Specifically, \texttt{pot} achieves the desired value of $\boldsymbol{u}$ with a rotation of the vector $(0,0,u)$ about $\boldsymbol{\hat{y}}$ by an angle $(\pi/2-\cos^{-1}c)$, followed by a rotation about $\boldsymbol{\hat{x}}$ by the randomly selected polar angle.

The position of reinjection can now be deduced; a position vector of length $r_b$ with zenith angle $\alpha$ and azimuthal angle $\psi$, where $\psi$ is sampled from a uniform distribution over $0$ and $2\pi$, is rotated about $\boldsymbol{u} \times \boldsymbol{\hat{z}}$ by an angle $\cos^{-1}(\boldsymbol{u} \cdot \boldsymbol{\hat{z}}/u)$. The motivation for this is that the randomly generated position vector would have the correct values of $\alpha$ and $b$ if $\boldsymbol{u}$ were parallel to $\boldsymbol{\hat{z}}$, and the rotation maps $\boldsymbol{\hat{z}}$ onto $\boldsymbol{\hat{u}}$ to generate the required injection position $\boldsymbol{r}$ for any given $\boldsymbol{u}$.

Finally \texttt{pot} computes the velocity $\boldsymbol{v}$ with which the particle enters the domain at $\boldsymbol{r}$ if its velocity at infinity is $\boldsymbol{u}$. This is done by assuming $\boldsymbol{u}$ is parallel to $\boldsymbol{\hat{z}}$ and using conservation of energy and angular momentum to deduce $v_r=-[u^2(1-b^2/r_b^2)-2q\phi(r_b)/m]^{1/2}$, $v_\varphi=0$ and $v_\theta=ub/r_b$. The rotation by angle $\cos^{-1}(\boldsymbol{u} \cdot \boldsymbol{\hat{z}}/u)$ around $\boldsymbol{u} \times \boldsymbol{\hat{z}}$ is then applied to give $\boldsymbol{v}$ in the general case where $\boldsymbol{u}$ is not parallel to $\boldsymbol{\hat{z}}$.

Several random numbers must be generated to execute the reinjection algorithm and to give the distribution of particles at the beginning of the simulation. Each process in a \texttt{pot} run generates its pseudorandom numbers with a WELL512 pseudorandom number generator \cite{Panneton2006, Lomont2011}, where each process uses its own 64-byte seed read from \texttt{/dev/urandom} (a special file provided on many Unix-like operating systems to produce pseudorandom bytes).

\section{Simulation results}
\label{sec:Results}
The plasma octree code \texttt{pot}, built to the specifications outlined in sections \ref{sec:Overview} and \ref{sec:Algorithms}, was tested to confirm its results' physical correctness. Sections \ref{sec:Validation} and \ref{sec:SOML} describe these tests.
In summary, these tests gave credible results, opening the way to investigating the effects of magnetization and grain non-sphericity with \texttt{pot}.
Section \ref{sec:HighB} presents \texttt{pot} results on the charging of a spherical dust grain in a magnetic field, while section \ref{sec:NonSpherical} presents more recent results on the charging of a non-spherical dust grain in unmagnetized plasma.
All of these simulations \tc{used} \texttt{pot}'s default values (table \ref{tab:defaults}), unless otherwise stated.

\tc{\texttt{pot} is typically run on 16 cores of Imperial College's CX1 cluster for several million timesteps, such that several microseconds elapse within each simulation.
On the $\mu$s timescale of the simulations the dust grain is essentially immobile.
Each simulation required a week to two months (depending on whether particle-particle interactions were calculated) of real time with the use of a realistic hydrogenic ion-to-electron mass ratio.
Results could be obtained more quickly were a smaller mass ratio used.}

\begin{table}
\begin{center}
\begin{tabular}{ l  c  c  c }
\toprule
description & flag & name & default value \\ \midrule
plasma particle count & \texttt{-N} & $N$ & \tc{150000} \\
time step & \texttt{-d} & $\delta t$ & $10^{-12}$ s \\
electron temperature & \texttt{-E} & $T_e$ & $220$ K \\
ion temperature & \texttt{-I} & $T_i$ & $220$ K \\
$\boldsymbol{\hat{x}}$ flow/drift speed & \texttt{-x} & $U$ & $0$ ms$^{-1}$ \\
simulation radius & \texttt{-m} & $R$ & $4 \times 10^{-4}$ m \\
sphere radius & \texttt{-a} & $a$ & $2.5 \times 10^{-6}$ m \\
\tc{aspect ratio} & \tc{\texttt{-A}} & \tc{$A$} & \tc{$1$} \\ \midrule
opening angle parameter &  & $\theta$ & 1 \\
ion-to-electron mass ratio &  & $m_i/m_e$ & 1836.15 \\
magnetic field &  & $\boldsymbol{B}$ & $(0,0,0)$ T \\
ion charge state &  & $Z$ & $+1$ \\
multipole expansion order &  &  & monopole \\ \bottomrule
\end{tabular}
\end{center}
\caption{The default parameter values of \texttt{pot}, which have been used to produce the results shown in this paper except in those cases where it is explicitly stated otherwise. The user may supply the first \tc{eight} parameters at run-time with the given flags, while adjusting the other parameters necessitates recompiling the program.}
\label{tab:defaults}
\end{table}

\subsection{Validation of core algorithms}
\label{sec:Validation}
The most basic test of the particle-motion integrator is to simulate a two-body system. Accordingly, \texttt{pot} has been used to simulate the very nearly circular Kepler orbit of an electron around a singly charged ion of mass $10^{15} m_e$. The electron's trajectory drifted by less than $0.001\%$ compared with its expected orbital path over $2 \times 10^9$ timesteps of length $10$ ps, confirming that the Boris integrator is suitable for the simulation.

\begin{figure}
\centering
\begin{floatrow}
\ffigbox[\FBwidth]
{
\subfloat{\includegraphics[width=0.44\textwidth]{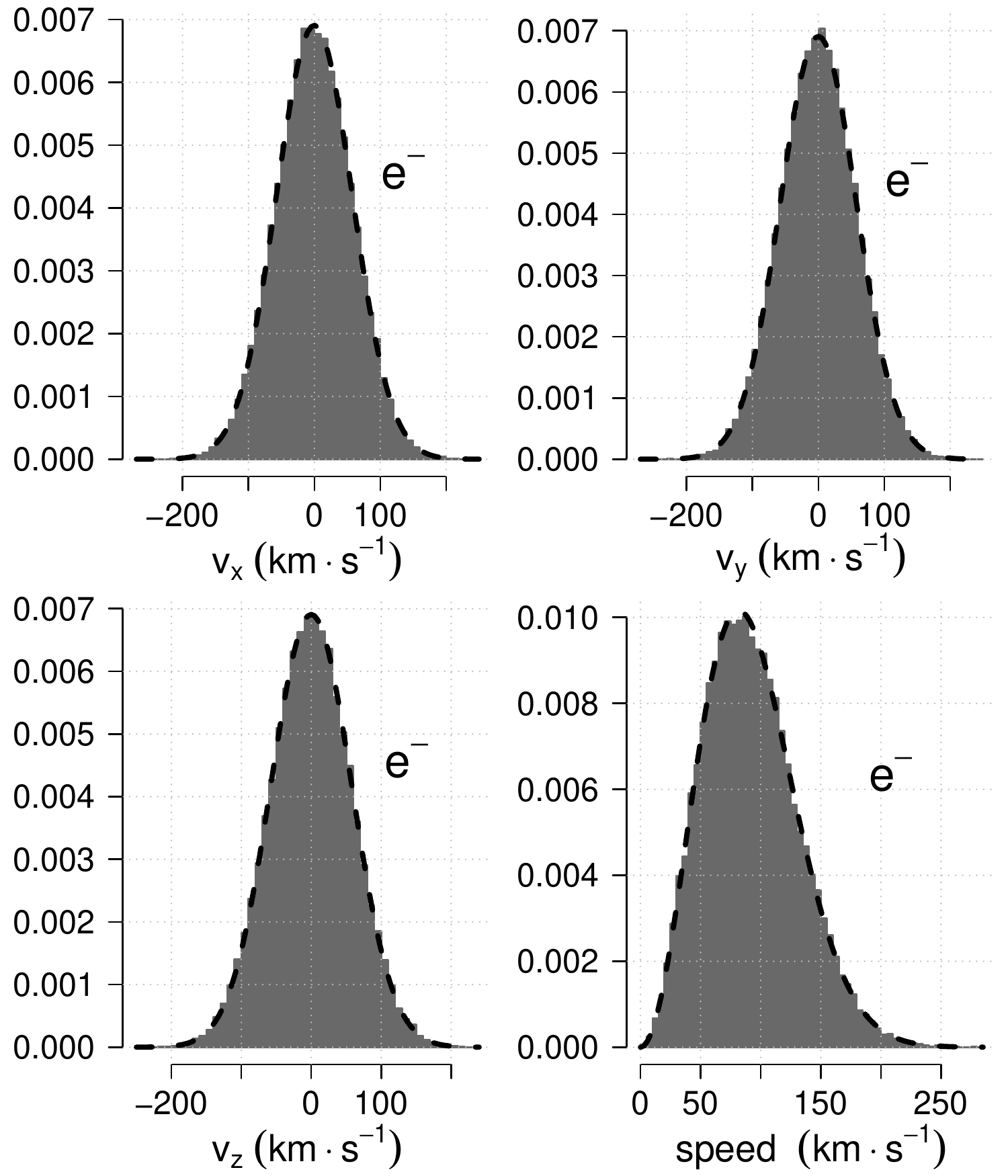}}
\subfloat{\includegraphics[width=0.44\textwidth]{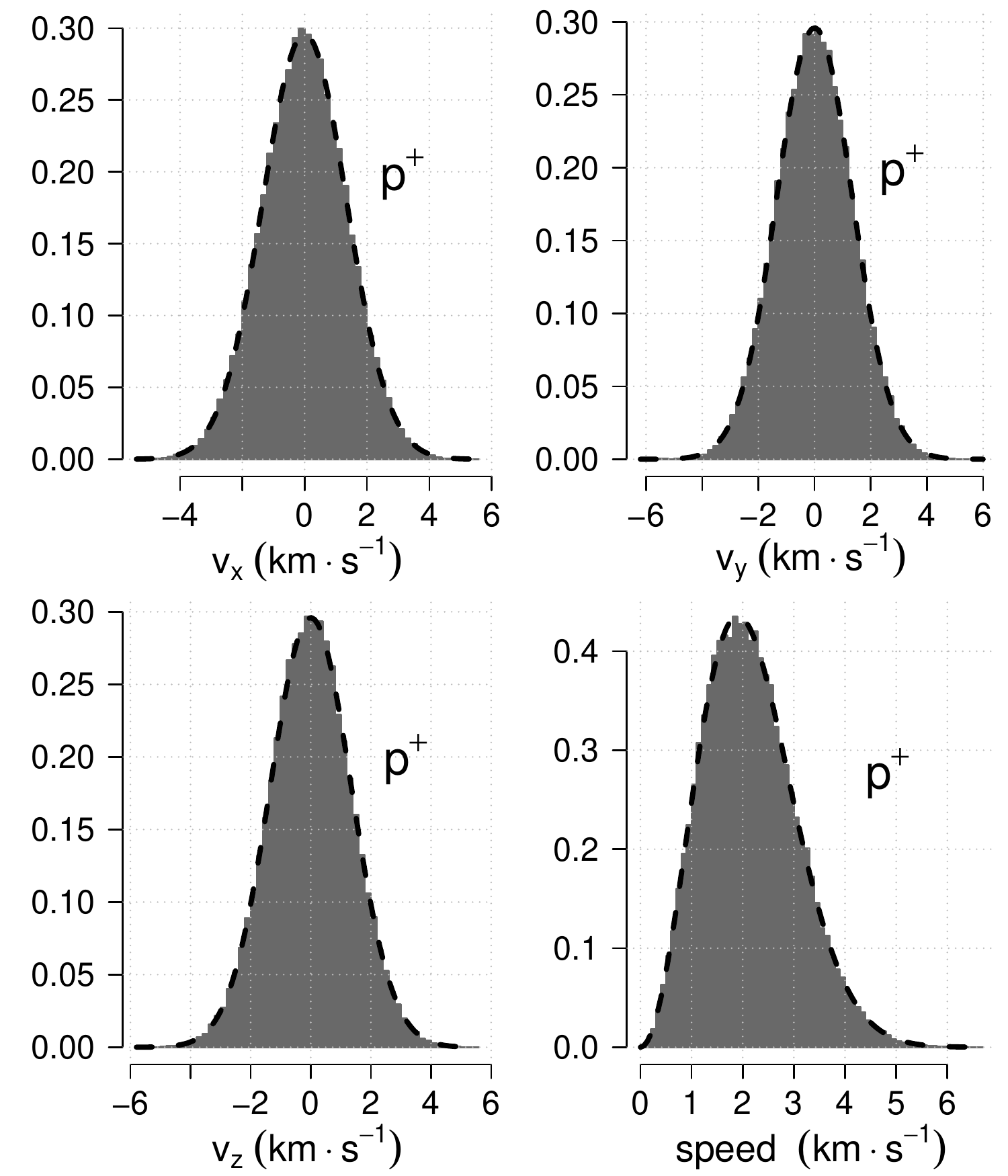}}
}
{\caption{Histograms of the electrons' and protons' speed and velocity components at equilibrium in a \texttt{pot} test run without a collecting sphere. The black, dashed curves represent the theoretical equilibrium distributions and excellent agreement is observed, indicating that the reinjection algorithm is operating as required.}\label{fig:Vdist}}
\end{floatrow}
\end{figure}

The particle-reinjection algorithm also requires validation. A test for this is to simulate a gas of non-interacting particles without a collecting sphere. The particles coast through the simulation domain in straight lines, placing minimal strain on the particle-motion integrator, so any variation in their mean energy or velocity distributions is solely due to the reinjection algorithm. \texttt{pot} has been run in this mode using its default values for a simulation of $6$ $\mu$s duration. The mean kinetic energy remained close to $(3/2)k_BT$ for both species as expected from kinetic theory. Figure \ref{fig:Vdist} shows histograms of the distributions of the velocity components and total velocity for each species at the end of the simulation, with the theoretically predicted Maxwell-Boltzmann distributions overlaid as black, dashed curves. Applying Anderson-Darling statistical tests \cite{Anderson-Darling} to the six velocity component distributions gives $p$-values over $0.5$ in all six cases, rigorously supporting the hypothesis that they are Gaussian as required. The implementation of Hutchinson's reinjection algorithm therefore passes this test with flying colours.

\begin{figure}[t]
{\includegraphics[width=0.9\textwidth]{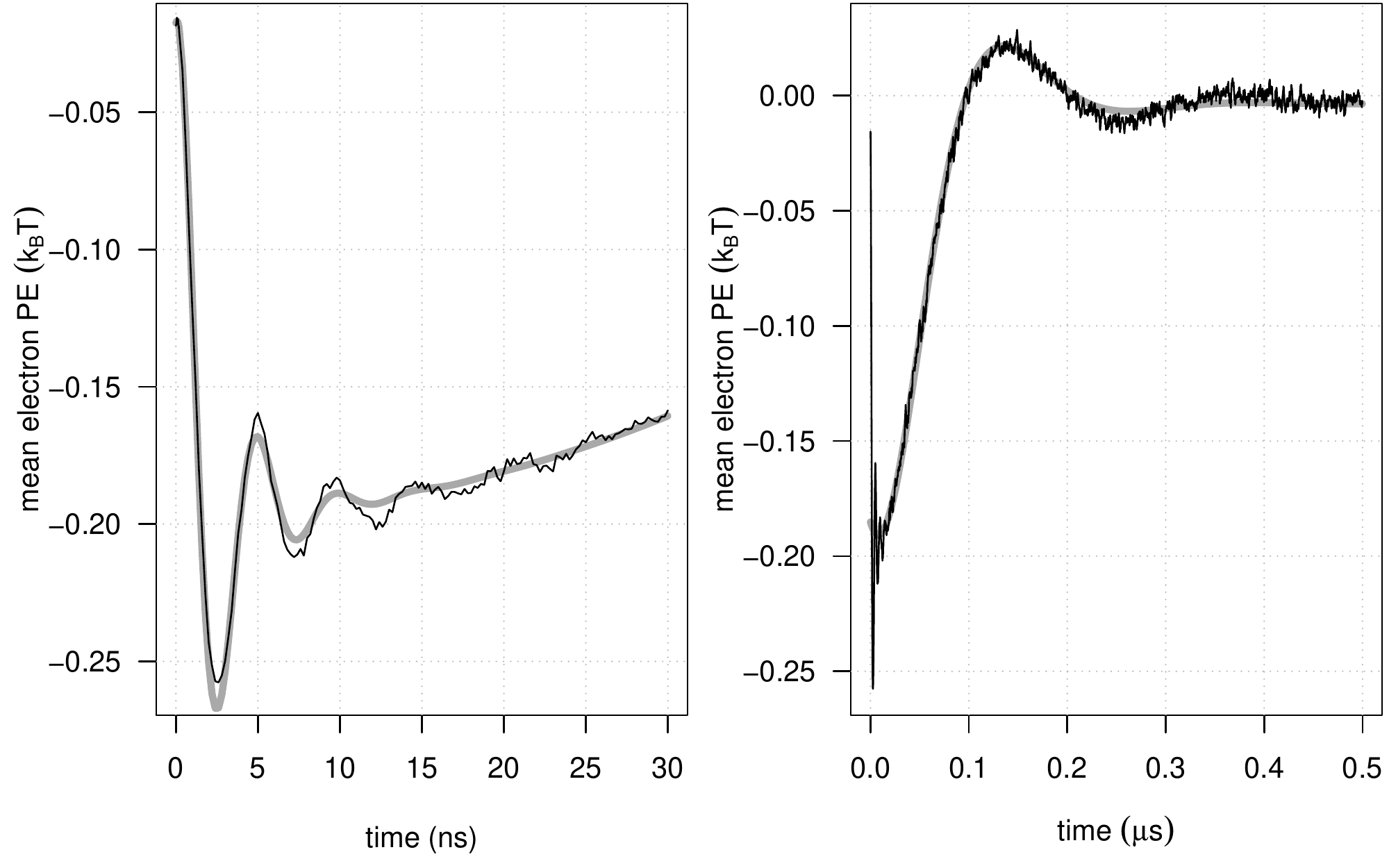}}
{\caption{The time evolution of mean electron potential energy for a \texttt{pot} simulation with non-equilibrium initial state. Oscillations are seen on the short timescale of electron plasma oscillations (left) and on the longer timescale of ion plasma oscillations (right). The overlaid non-linear oscillation curves, from equation (\ref{eqn:fitequation}) with the least-squares-fitted parameters shown in table \ref{tab:oscillationfits}, are shown as thick grey lines.}\label{fig:PlasmaOsc}}
\end{figure}

Finally the treecode must be tested to check whether it accurately calculates particle interactions. The treecode must be able to predict collective phenomena arising from these interactions, so \texttt{pot} is run without a collecting sphere but with interacting particles. The simulation is initialized in a non-equilibrium state, by distributing the initial distances of particles from the simulation centre according to the square root of uniformly sampled variates rather than the correct cube root distribution, to see whether plasma oscillations are reproduced. The simulation therefore begins with an excess of particles near the centre. The electrons, having higher thermal velocity than the ions, are expected to rush into the low-density region near the simulation boundary ahead of the ions. This sets up a separation of charge, pulling the electrons back towards the centre, causing the particles to undergo damped plasma oscillations until an equilibrium state is reached. This oscillatory behaviour is indeed seen in figure \ref{fig:PlasmaOsc}, which shows the variation in the mean electron potential energy, $V(t)$, from the start of the simulation. Two modes of oscillations, on different timescales, are seen which correspond to fast electron and slow ion oscillations. Oscillation parameters were extracted from these results by fitting the formula for exponentially decaying oscillations,
\begin{equation}
 V(t) = V_a \exp \left( \frac{-t}{\tau_d} \right) \cos( \omega t + \varphi) + C + \left[ \frac{t^2}{\tau_s^2} \right],
\label{eqn:fitequation}
\end{equation}
where $V_a$ is the oscillation amplitude, $\tau_d$ is the decay time, $\omega$ is the oscillation frequency, and $\varphi$ and $C$ are constants for initial phase and offset. The term in square brackets is an optional quadratic time trend, with time scale $\tau_s$, which is included for the fast oscillation to account for the slow oscillation superimposed on it.
Table \ref{tab:oscillationfits} summarizes the estimates of equation (\ref{eqn:fitequation})'s parameters and figure \ref{fig:PlasmaOsc} includes the corresponding curves.
\texttt{pot} reproduces plasma oscillations with frequencies similar to those expected, evidence that its treecode-algorithm implementation functions properly.

\begin{table}[!h]
\begin{center}
\begin{tabular}{lr@{ }c@{ }lr@{ }c@{ }l}
\toprule
parameter & \multicolumn{6}{c}{estimate} \\
\cmidrule{2-7}
& \multicolumn{3}{c}{fast oscillation}
& \multicolumn{3}{c}{slow oscillation} \\
\midrule
$C$ $(k_B T)$ & $-0.1970$ & $\pm$ & 0.0003 & $-0.0035$ & $\pm$ & 0.0001 \\
$\tau_s$ (ns)    & $-157$    & $\pm$ & 2      & \multicolumn{3}{c}{omitted} \\
$V_a$ $(k_B T)$ & $0.1925$   & $\pm$ & 0.0007  & $0.252$ & $\pm$ & 0.001 \\
$\tau_d$ (ns)    & $2.61$    & $\pm$ & 0.03   & $63.3$ & $\pm$ & 0.3 \\
$\omega$ (rad$\cdot$ns$^{-1}$) & $1.293$ & $\pm$ & 0.003 & $0.0247$ & $\pm$ & 0.0001 \\
phase (rad)      & $-0.37$   & $\pm$ & 0.01   & $2.374$ & $\pm$ & 0.007 \\
\midrule
$\omega_{pe}$ and $\omega_{pi}$ \big(rad $\cdot$ ns$^{-1}$\big)
& 0.94 & & & 0.0220 & & \\
\bottomrule
\end{tabular}
\caption{Estimates of the parameters in equation (\ref{eqn:fitequation}) for the fast and slow oscillations shown in figure \ref{fig:PlasmaOsc}. The last two lines give the undamped, linear-oscillation plasma frequencies $\omega_{pi}$ and $\omega_{pe}$.}
\label{tab:oscillationfits}
\end{center}
\end{table}

\subsection{Validation against SOML theory}
\label{sec:SOML}
The previous section's tests, although encouraging, neglected the presence of a collecting sphere. The purpose of \texttt{pot} is to simulate a plasma in the vicinity of such an object, so \texttt{pot} simulated a flowing plasma with inter-particle interactions and a collecting sphere at the centre of the simulation domain.
The SOML theory \cite{Willis2012} provides a benchmark \tc{by} describ\tc{ing} the charging of a sphere under \tc{\texttt{pot}'s default} conditions \tc{where the dust grain radius is much smaller than the Debye length}.
\tc{The charging of larger dust grains, where $a$ becomes comparable to the Debye length, is deferred to later studies.}
In contrast to the tests of the previous section, a disagreement between \texttt{pot} and the theory here would not necessarily be a reproach to \texttt{pot} as discrepancies might instead be attributable to a violation of the SOML theory's assumptions.

\begin{figure}
\floatbox[{\capbeside\thisfloatsetup{capbesideposition={right, center},capbesidewidth=0.48\textwidth}}]{figure}[\FBwidth]
{\caption{The charge $q$, in units of $e$, of an absorbing sphere for a \texttt{pot} run with no external fields or plasma flows. The OML-theory prediction \cite{Allen1992} is overlaid as a thick dashed line. The region where the charge fluctuations are within $1\sigma$ of the equilibrium value, with $\sigma$ provided by stochastic modelling \cite{Thomas2013}, is shaded grey. Both theoretical predictions agree well with \texttt{pot}'s results.}\label{fig:OMLcharging}}
{\includegraphics[width=0.48\textwidth]{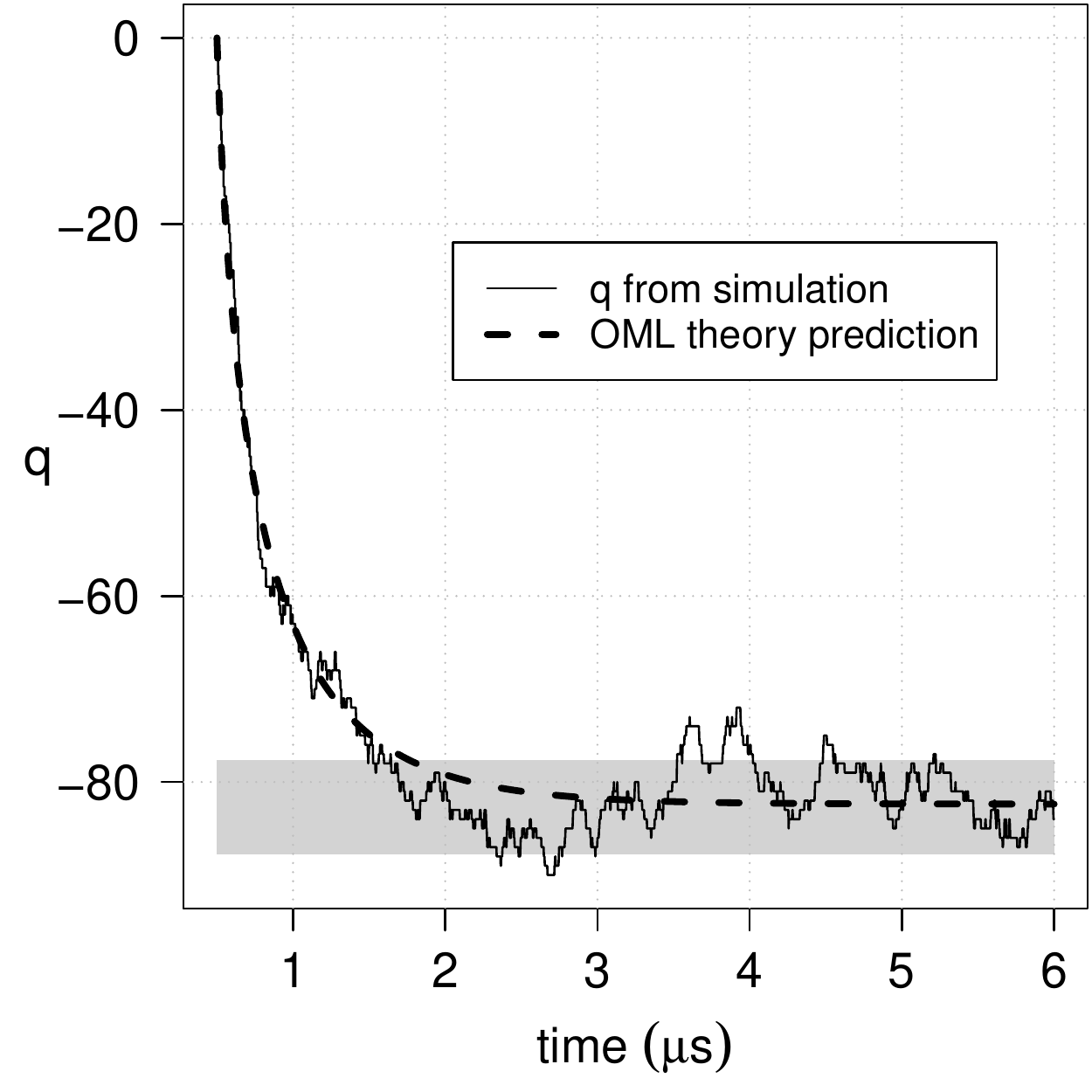}}
\end{figure}

The specific case of zero plasma flow velocity is considered first. The central sphere is initialized with no charge and allowed to collect charge from the plasma; the evolution of this charge over the course of the simulation is plotted in figure \ref{fig:OMLcharging}. The charge evolution predicted by OML theory \cite{Allen1992} has been added to the plot along with a shaded region to represent the charge's standard deviation $\sigma$ predicted by stochastic modelling \cite{Thomas2013}. The charge of the sphere as calculated by \texttt{pot} is consistent with both of these theories; applying a $\chi^2$ goodness-of-fit test to the variance $\sigma^2$ from the stochastic model and \texttt{pot}'s results gives a $p$-value of $0.66$, indicating statistically insignificant disagreement.

\texttt{pot} has been run under these conditions seven additional times and the median charge calculated for the period after $5$ $\mu$s, after which point all the simulations had equilibrated, for each run. This charge can be converted to the sphere's surface potential, $\phi_a$, by dividing by the capacitance of a conducting sphere, $4\pi \epsilon_0 a$. The normalized surface potential of the sphere, $\eta_a=e\phi_a/(k_BT_e)$, has a mean and standard deviation of $-2.56$ and $0.070$ respectively across all eight runs. This mean is $2\%$ less than the OML-predicted value of $-2.50$, albeit with a random error of $3\%$. Similar discrepancies of around $2\%$ have previously been reported between the OML theory and PIC codes; this effect has been tentatively attributed to the development of an absorption barrier for the ions \cite{Delzanno2013}.

\begin{figure}
\floatbox[{\capbeside\thisfloatsetup{capbesideposition={right, center},capbesidewidth=0.46\textwidth}}]{figure}[\FBwidth]
{\caption{\texttt{pot}'s median equilibrium $\eta_a$ as a function of plasma flow speed, versus SOML theory's $\eta_a$ predictions \cite{Willis2012}, for $\Theta = 1$ (solid curve) and $\Theta = 0.1$ (dashed curve). The \texttt{pot} results follow the SOML trends, although the former show a systematic overcharging by $2\%$ and a random error of $3\%$.}\label{fig:SOMLcomparison}}
{\includegraphics[width=0.48\textwidth]{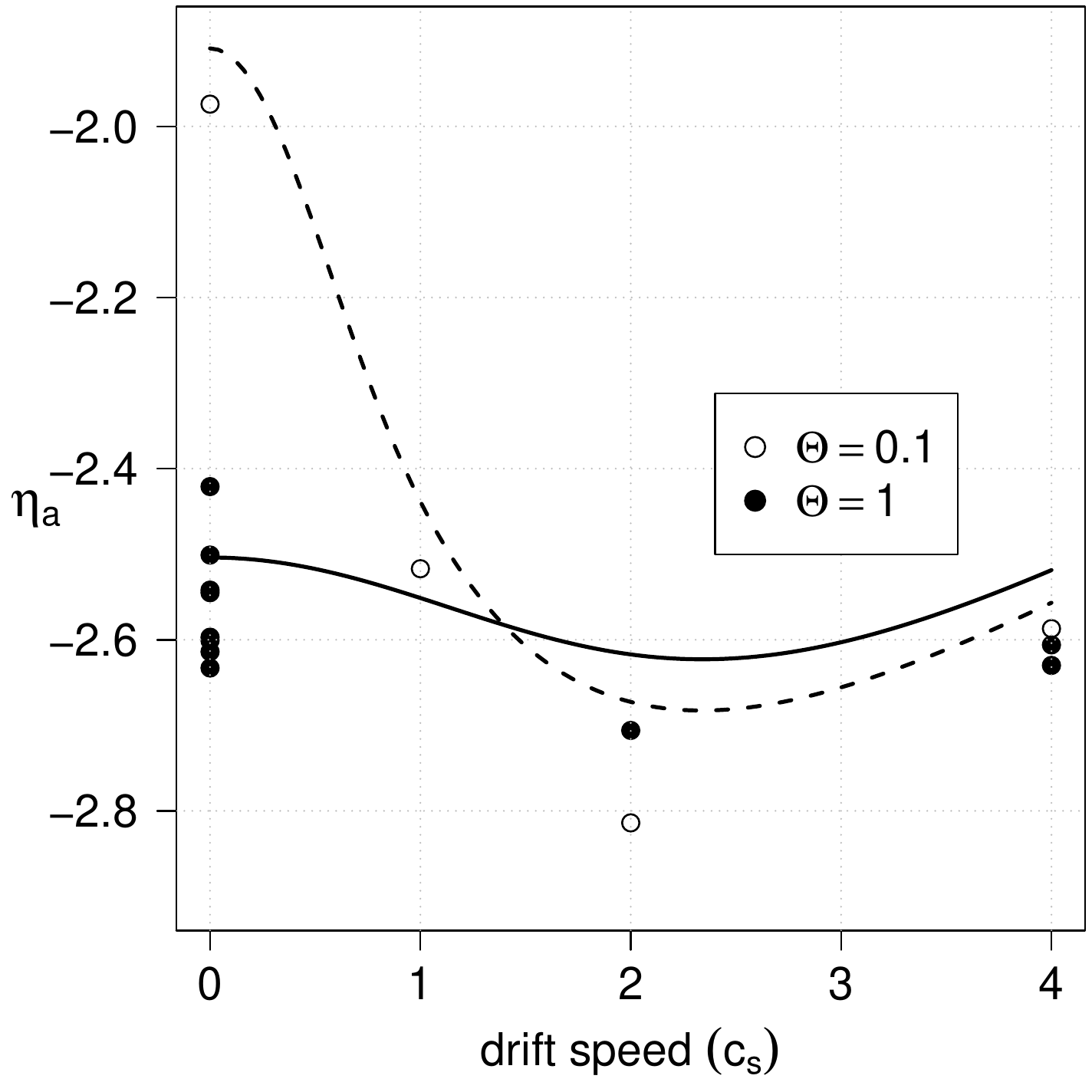}}
\end{figure}

We now consider the more general case of flowing plasmas.
Figure \ref{fig:SOMLcomparison} plots the equilibrium floating potential's median value against the drift speed of the plasma, along with curves representing SOML theory's predictions.
So far the ion-to-electron temperature ratio, $\Theta \equiv T_i / T_e$, has been $1$; we now add the case of $\Theta = 0.1$.
While \texttt{pot}'s results follow the general trends of the SOML model, the calculated floating potential is systematically more negative than the SOML value.
This difference may be attributable to \texttt{pot}'s inclusion of Coulomb collisions or the fact the SOML is only strictly valid for vanishingly small dust grains.

\subsection{Charging of spherical dust in magnetic fields}
\label{sec:HighB}
The behaviour of dust in magnetized plasmas is of vital importance to studies of dust transport in magnetic fusion devices \cite{Bacharis2012} and alignment of dust grains in the interstellar medium \cite{Jones1967}. Recent results from \texttt{pot} summarize the calculation of the normalized floating potential $\eta_a$ of a collecting sphere in a static, homogeneous magnetic field. This has been a contentious area of recent research, as section \ref{sec:Intro} alluded to, following Tsytovich \textit{et al.}'s hypothesis of a regime where electrons are fully magnetized but ions unmagnetized on the length scale of the dust grain \cite{Tsytovich2003}. SCEPTIC has already been applied to test this hypothesis and finds that it holds only at very weak magnetic fields \cite{Patacchini2007}. However the floating potentials computed by SCEPTIC must be subject to some doubt because SCEPTIC assumes a Boltzmann relation for the electrons, which may be invalid in the presence of a magnetic field \cite{Allen2008}. \texttt{pot} makes no dubious assumptions regarding the Boltzmann relation and, as a fully microscopic simulator, is ideal for testing Tsytovich's hypothesis.

\begin{figure}
\centering
\begin{floatrow}
\ffigbox[\FBwidth]
{
\subfloat{\includegraphics[width=0.44\textwidth]{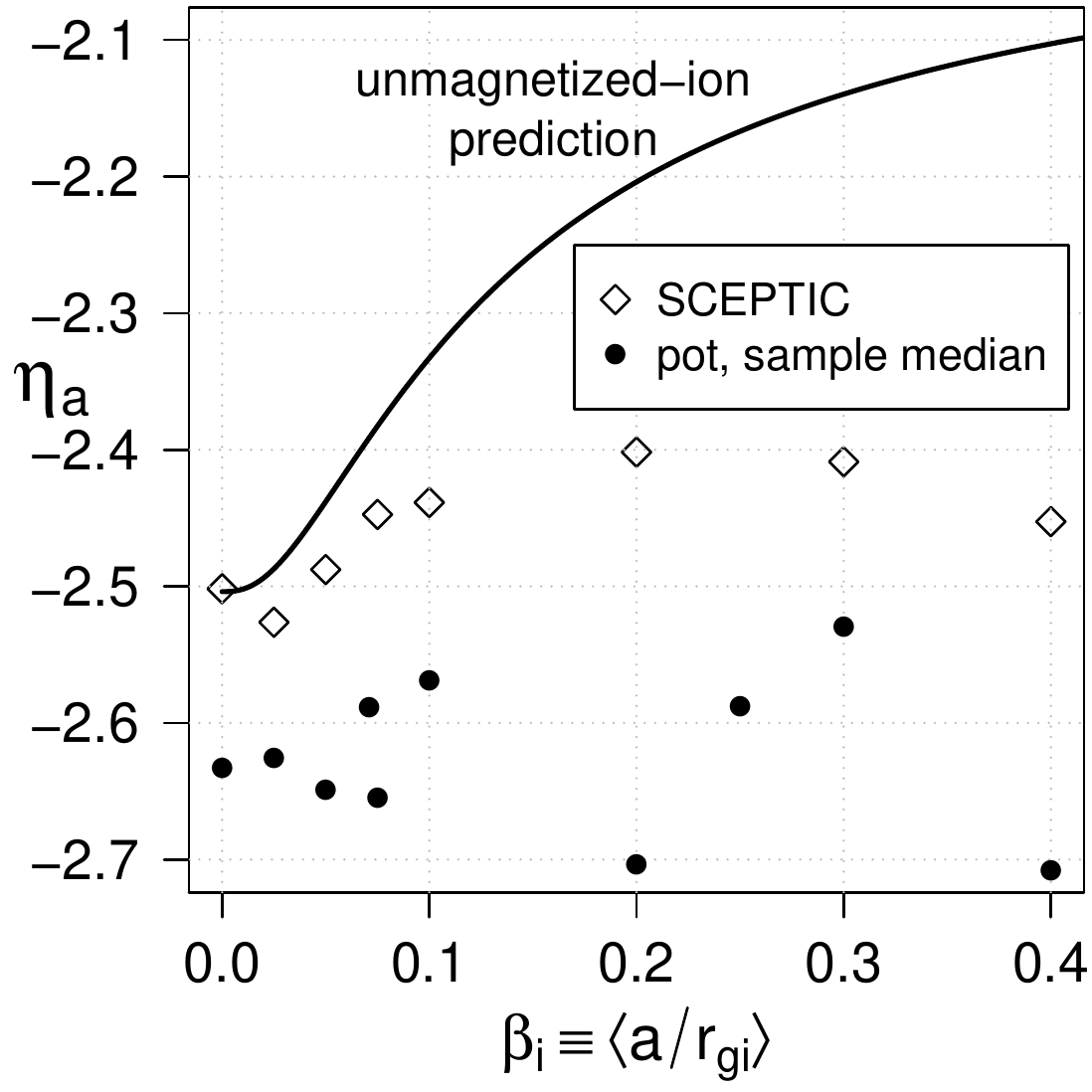}}
\subfloat{\includegraphics[width=0.44\textwidth]{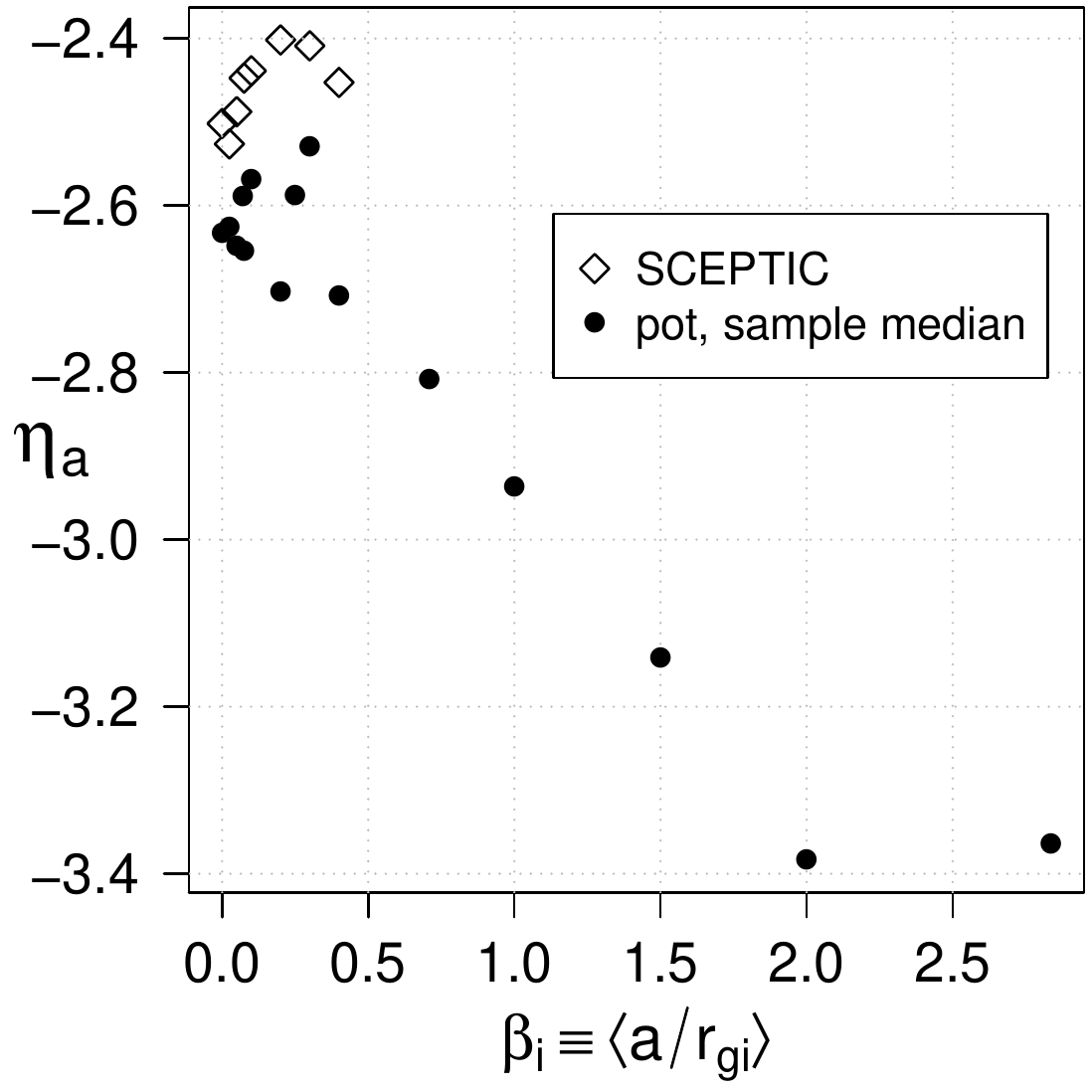}}
}
{\caption{$\eta_a$ as a function of $\beta_i$, as estimated using: SCEPTIC \cite{Patacchini2009}, median equilibrium values in \texttt{pot} runs, and the assumption that ions remain unmagnetized \cite{Tsytovich2003}. \texttt{pot}'s results support rejection of the unmagnetized-ion hypothesis. \texttt{pot} does not assume the Boltzmann relation for electrons, so its results can be reliably extended to high magnetic field strengths.}\label{fig:Bcharging}}
\end{floatrow}
\end{figure}

Figure \ref{fig:Bcharging} shows results for $\eta_a$, obtained as the median of the equilibrated surface potential, of a sphere in a stationary plasma (with \texttt{pot}'s default parameters) plotted against varying magnetic field strength. The field strength is parameterized by the mean ratio of the dust grain radius to the ion gyroradius
\begin{equation}
 \beta_i \equiv \left< \frac{a}{r_{gi}} \right> = a \left( \frac{\pi k_BT_im_i}{2Z^2e^2B^2} \right)^{-\frac{1}{2}}
 \label{eq:beta-i-defn}
\end{equation}
and the magnetic field is always in the $\boldsymbol{\hat{x}}$-direction such that $\boldsymbol{B}=B\boldsymbol{\hat{x}}$; this still represents an arbitrary direction due to the spherical symmetry of the system. The results obtained from SCEPTIC and the unmagnetized-ion theory have been added to the left panel and show that \texttt{pot}'s results support the rejection of the unmagnetized-ion theory. The results of \texttt{pot} and SCEPTIC both show $\eta_a$ depending only weekly on $\beta_i$ when $\beta_i$ is small, but \texttt{pot} consistently gives floating potentials $7\%$ more negative than SCEPTIC. This is much larger than the $2\%$ systematic offset between \texttt{pot} and the SOML theory. The probable reason for this is suggested by comparison with the recent publication, earlier this year, of new PIC-code results with fully simulated, rather than Boltzmann, electrons \cite{Lange2016}. That work also calculates the dust grain's floating potential as being more negative than SCEPTIC; its main text states a $5\%$ difference but the graphically presented results suggest it is slightly higher. Although the PIC code is collisionless and calculates macroscopic electric fields only, it suggests that discrepancies between \texttt{pot} and SCEPTIC are due to SCEPTIC's incorrect assumption that the Boltzmann relation for electrons is valid in a magnetic field.

The treecode results have been extended to higher magnetic fields as shown by the right panel of figure \ref{fig:Bcharging}. These results show the floating potential of the grain tending towards the fully-magnetized-ion-and-electron limit of $\eta_a = -3.76$ and corroborate the PIC-code results in \cite{Lange2016}.

\subsection{Charging of non-spherical dust grains}
\label{sec:NonSpherical}
In spite of natural dust grains having a wide variety of shapes, almost all existing theories of dust grain charging, including those considered in the preceding sections of this paper, address only spherical grains.
Because it is important to know how the charging of non-spherical grains differs from that of spheres, Holgate and Coppins recently extended OML theory to calculate the floating potentials of spheroids and quantify the effect of a grain's oblateness on its charge \cite{Holgate2016}.
\texttt{pot} is an ideal code to test this spheroidal-OML theory, because \texttt{pot}'s treecode algorithm is a mesh-free computational method which can accommodate simulation domains with complicated geometries without needing cumbersome changes to its remeshing algorithm.

The source code of \texttt{pot} includes options to compile \texttt{pot} with either an oblate or prolate spheroid in place of the central sphere.
In these cases the analytical vacuum solution of a conducting spheroid gives the grain's electric field, and the algorithm for collection of electrons and ions is modified to account for the grain's spheroidal surface.
A simulation runs in the same basic manner as in the spherical case: the initially uncharged grain collects electrons and ions from the plasma until obtaining its equilibrium charge.

The simulations in this subsection ran with $T_e = T_i = 1$ eV to reduce the effect of fluctuations on the equilibrium charge value.
In itself, however, the higher temperature would have increased the Debye length, so the simulation domain's radius $R$ was simultaneously cut to 262 $\mu$m, increasing the plasma density and ensuring the Debye length did not exceed $R$.
All of the other simulation parameters retained their default values (table \ref{tab:defaults}).
A timestep of 1 ps continued to satisfy the requirements section \ref{sec:Boris} details.

\begin{figure}
\floatbox[{\capbeside\thisfloatsetup{capbesideposition={right, center},capbesidewidth=0.46\textwidth}}]{figure}[\FBwidth]
{\caption{A comparison of the floating potentials of spheroidal dust grains in a hydrogen plasma according to the spheroidal-OML theory \cite{Holgate2016}, represented as a curve, and the mean equilibrium values of \texttt{pot} simulations, represented by circles with standard deviations of $\eta_a$ shown by error bars. When the systematic offset is taken into account \texttt{pot} agrees well with OML theory for $A \sim 1$, but \texttt{pot} predicts that highly deformed spheroids deviate further from the spherical value of $\eta_a$ than the spheroidal-OML theory predicts.}\label{fig:nonspherical}}
{\includegraphics[trim={1cm 0.5cm 0.3cm 0.8cm},clip,width=0.45\textwidth]{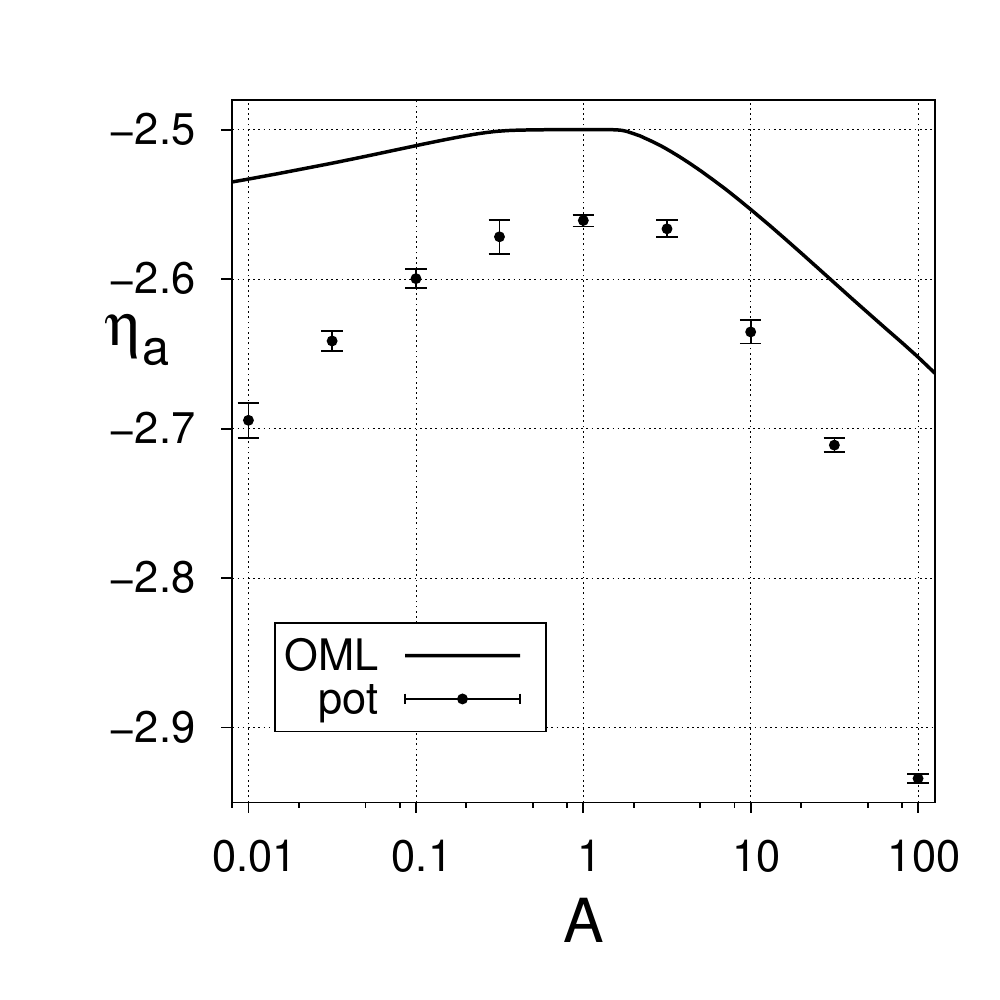}}
\end{figure}

Figure \ref{fig:nonspherical} compares the \texttt{pot} results for the equilibrium floating potential of spheroids to the spheroidal-OML theory for a range of aspect ratios, $A$, with the volume of the spheroids kept constant.
The aspect ratio is the ratio of the lengths of the symmetric axis of the spheroid to the equatorial axes of the spheroid; elongated prolate spheroids therefore have $A > 1$ while flattened oblate spheroids have $A < 1$.
Both \texttt{pot} and the spheroidal-OML theory show that spheroids have floating potentials of slightly larger magnitude than spheres, and in absolute terms the disagreement between \texttt{pot} and theory is modest.
This offset is not affected by the temperature and density of the particles in the simulation, nor by the size of the simulation domain.
However the offset worsens for highly deformed spheroids; the probable cause is inadequate resolution of particles' motion near the sharp edges of the spheroids.
The tips of a prolate spheroid have radii of curvature $aA^{-4/3}$, where $a$ is the radius of a sphere with the same volume, which is $0.002a$ when $A=100$.
Such small effective radii can lead to violation of condition (i) in section \ref{sec:Boris}; a shorter simulation timestep should provide more accurate results at the expense of longer runtimes.

\section{Summary}
\label{sec:Summary}

The plasma octree code \texttt{pot} has been developed to examine the validity of prevailing theories of dust-plasma interactions and to predict new dust-plasma behaviours. As a fully microscopic simulation of a plasma in the vicinity of a dust grain, this treecode has advantages over the  methods currently used; for example, it does not assume the Boltzmann relation for electrons, and Coulomb collisions between particles are inherent and are not artificially imposed. As a mesh-free code, \texttt{pot} can also handle non-symmetric simulation domains with relative ease. \texttt{pot} has been tested against existing theories and simulations; this mutually verifies not only the accuracy of \texttt{pot}, but also the validity of the assumptions made in these existing works. The results obtained thus far support the widely used OML and SOML charging theories, and the more recent spheroidal-OML theory, but call into question the validity of using a Boltzmann relation in hybrid PIC codes, particularly in the presence of a magnetic field.

\texttt{pot} employs several noteworthy algorithms. It provides the first implementation of the Barnes-Hut treecode algorithm in a low-temperature plasma environment and represents the first time that Hutchinson's particle-reinjection algorithm has been used outside SCEPTIC. It is also unusual in its use of the Boris particle-motion integrator outside a particle-in-cell context. A review of all three algorithms has been provided for the benefit of researchers wishing to understand the operation of \texttt{pot} or develop their own treecode simulation.

The treecode method can be used to model various aspects of dust grains in a plasma beyond those discussed in this paper; examples include the drag force exerted by the plasma on the dust grains \cite{Patacchini2008}, the torque applied to the dust grains by the plasma \cite{Tsytovich2003}, and the interactions between two or more dust grains \cite{Lampe2015}. As such the treecode method, and its implementation in the plasma octree code \texttt{pot}, could well become a vital tool in the future study of dusty plasmas.

\section*{Acknowledgements}
We would like to thank both our supervisor, Dr.\ Michael Coppins, and Prof.\ John Allen for their insights and support. The simulations reported on in this paper were run on the Imperial College High Performance Computing Service's CX1 cluster. This work has been supported by the UK's Engineering and Physical Sciences Research Council.\\

%% bibliography

\end{document}